# Dynamic Cognitive Process Application of Blooms Taxonomy for Complex Software Design in the Cognitive Domain

**Shashi Kumar NR [1], Pushpavathi TP[2], Prof. Selvarani R[3]**

[1] 2nd Year M.Tech. Student Dayanand Sagar College of Engineering, Bangalore. email: nr.shashikumar@yahoo.com
[2] 2nd Year M.Tech. Student Dayanand Sagar College of Engineering, Bangalore. email: acepushpa@yahoo.co.in
[3] Professor , Dayanand Sagar College of Engineering, and Head, SERIG,RIIC,DSI,Bangalore .email: selvss@yahoo.com

(Contact person's Address)
Prof. Selvarani R
Department of Computer Science & Engineering,
Dayanand Sagar College of Engineering,
Kumaraswamy Layout
Bangalore-560078, India
Email: selvss@yahoo.com

**ABSTRACT**

Software design in Software Engineering is a critical and dynamic cognitive process. Accurate and flawless system design will lead to fast coding and early completion of a software project. Blooms taxonomy classifies cognitive domain into six dynamic levels such as Knowledge at base level to Comprehension, Application, Analysis, Synthesis and Evaluation at the highest level in the order of increasing complexity. A case study indicated in this paper is a gira system, which is a gprs based Intranet Remote Administration which monitors and controls the intranet from a mobile device. This paper investigates from this case study that the System Design stage in Software Engineering uses all the six levels of Blooms Taxonomy. The application of the highest levels of Blooms Taxonomy such as Synthesis and Evaluation in the design of gira indicates that Software Design in Software Development Life Cycle is a complex and critical cognitive process.

# Dynamic Cognitive Process : Application of Bloom's Taxonomy for Complex Software Design in the Cognitive Domain

## INTRODUCTION

Computer software has become a driving force. It is the engine that drives business decision making. It serves as the basis for modern scientific investigation and engineering problem solving. Software is virtually inescapable in a modern world. Software's impact on our society and culture continues to be profound. As its importance grows, the software community continually attempts to develop technologies that will make it easier, faster and less expensive to build high-quality computer programs[6].

Systems are created to solve problems. One can think of the systems approach as an organized way of dealing with a problem. The objectives of Systems Analysis and Design are to understand a system, know the components of system. A collection of components that work together to realize some objective forms a system. Basically there are three major components in every system, namely input, processing and output as shown in the Figure1.

**Fig 1 : A System Components**

The different phases of software development life cycle are System study , Feasibility study , System analysis, System design, Coding , Testing , Implementation , Maintenance. The different phases of software development life cycle are as shown in Fig. 2.

Figure2: Software Development Cycle

Based on the user requirements and detailed analysis, a new system must be designed. It is a most crucial phase in the development of a system. Normally, the design proceeds in two stages, they are Preliminary or general design and Structured or detailed design.

In the preliminary or general design, the features of the new system are specified. The costs of implementing these features and the benefits to be derived are estimated. If the project is still considered to be feasible, we move to the detailed design stage.

In the detailed design stage, computer oriented work begins in earnest. At this stage, the design of the system becomes more structured. Design is a blue print of a computer system solution which establishes an interrelationship among the same components as the original problem. Input, output and processing specification are drawn up in detail. In the design stage, the programming language and the platform in which the new system will run are also decided.

Software design is actually a multi-step process that focuses on four distinct attributes of a program: data structure, software architecture, interface representations and algorithmic detail.[6]
Software design sits at the technical kernel of software engineering. Beginning once software requirements have been analyzed and specified, software design is the first of three technical activities – design, code generation and test that are required to build and verify the software.[6]
There are several tools and techniques used for designing. These tools and techniques are Flowchart, Data flow diagram, Data dictionary, Structured English, Decision table, Decision tree etc.

## Cognitive Informatics

Cognitve Informatics (CI) is an emerging discipline that studies the natural intelligence and internal information processing mechanisms of the brain, as well as the processes involved in perception and cognition. CI provides a coherent set of fundamental theories and contemporary mathematics which form the foundation for most information and knowledge based science and engineering disciplines such as computer science, neuropsychology, software engineering.[1]
Cognitive Informatics treats the natural world as a triple of <I, E, M>, where I is Information, E energy and M Matter. Information is used to model the abstract world, whereas enery and matter are used to model the physical world. [16]
Learning is the cognitive process that obtains the required knowledge. The difficulty of learning is classified by loom's Taxonomy. Bloom identified three domains of learning : cognitive, affective and psychomotor.[4]. The cognitive domain is for mental skills, the affective domain is for growth in feelings or emotional areas and the psychomotor domain is for manual or physical skills.

## Bloom's Taxonomy

In 1956, Benjamin Bloom headed a group of educational psychologists who developed a classification of levels of intellectual behavior important in learning. Bloom found that over 95 % of the test questions students encounter require them to think only at the lowest possible level...the recall of information.
Bloom identified six levels within the cognitive domain, from the simple recall or recognition of facts, as the lowest level, through increasingly more complex and abstract mental levels, to the highest order which is classified as evaluation. The figure 3 below shows the pyramid of Bloom's taxonomy.

Verb examples that represent intellectual activity on each level are listed below:

.***Knowledge***: arrange, define, duplicate, label, list, memorize, name, order, recognize, relate, recall, repeat, reproduce state.

.***Comprehension***: classify, describe, discuss, explain, express, identify, indicate, locate, recognize, report, restate, review, select, translate.

.***Application***: apply, choose, demonstrate, dramatize, employ, illustrate, interpret, operate, practice, schedule, sketch, solve, use, write.

.***Analysis***: analyze, appraise, calculate, categorize, compare, contrast, criticize, differentiate,

discriminate, distinguish, examine, experiment, question, test.

. ***Synthesis***: arrange, assemble, collect, compose, construct, create, design, develop, formulate, manage, organize, plan, prepare, propose, set up, write.

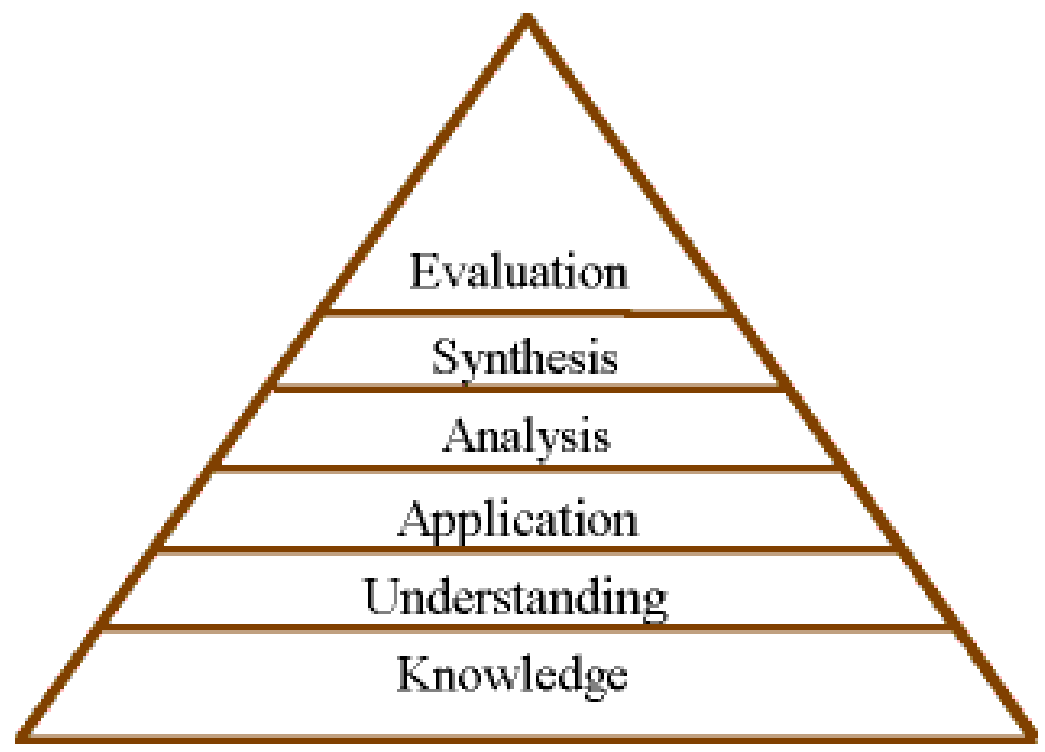

Figure 3. Pyramid of Bloom's Taxonomy

.***Evaluation***: appraise, argue, assess, attach, choose compare, defend estimate, judge, predict, rate, core, select, support, value, evaluate.

**Case Study**

We conducted a case study in order to understand the cognitive activities involved in software design process and its relationship to Bloom's Taxonomy. This is an observational case study. In this case study, we have taken a software project development titled ' GPRS based Intranet Remote Administration' [GIRA].

**GIRA**

In a world of increasing mobility, there is a growing need for people to communicate with each other and have timely access to information regardless of the location of the individuals or the information. With the advent of new technology, the way of communication is also changed. Mobile phones are long range, portable and wireless electronic device of communication.

The GIRA system is basically a mobile phone service. In this paper we are discussing about a novel local area network control system called 'GPRS based Intranet Remote Administration' (GIRA). This system finds application in the common portable device. With this system, a network administrator will have an effective remote control over the network. GIRA system is developed using GPRS, GCF (Generic Connection Framework of J2ME), Sockets and RMI Technologies.

This GIRA system will enable a Network Administrator to monitor and control the intranet thru a mobile device. Remote Administration of intranet in terms of obtaining the list of client PCs, to run or abort a process in a specific client PC or a server, to execute certain applications, chat with a user or can broadcast a message and also will be able to shut down a client or a server remotely. The fig. 4 shows the Main Flowchart.

**1. Client List:** An administrator will be able to view the names of all the clients connected to the server along with the clients.

**2. Process:** An administrator will be able to choose any one of the clients among the given

list, and he can choose the options like1) List of process
2) Start the process3) Kill (abort) the process.
**3. Compile:** An administrator can compile a file he wishes to do from a mobile device after specifying the path and file name.
**4. WordPad:** A user can type a text and save into a file of a particular client.
**5. Read:** An user will be provided with all the drives existing in a specific client and any file of a particular drive can be opened for read in a mobile device.
**6. Execute:** This menu facilitates a user from compiling any java programs from a mobile device after explicitly mentioning a particular java file that exists in a particular drive of a specific client.
**7. Chat:** This option enables the administrator to have on line interaction with any of his clients.
**8. Broadcast:** This option facilitates an administrator for broadcasting a message to all the clients and a server from a mobile device.
**9. Client Messaging:** This option provides the facility to the administrator to send messages to a particular client.

**10. Shut Down:** This option closes the entire operation.

## Bloom's Taxonomy in GIRA's Design

We are here trying to match the activities of the software design with that of the six levels of Bloom's Taxonomy We shall examine the application of all the levels in detail with respect to the design of GIRA.

**Knowledge** is the first and base level of Bloom's Taxonomy, the keywords associated with this level is recall, identify, recognize, define, describe and reproduce. This is about recalling information, knowing the prerequisites to understand the multi-tier architecture, to know the client-server system, java programming, mobile programming, GPRS technology, RMI, socket programming concepts, HTTP and network programming. It is also necessary to know and understand the full requirement of how to implement the idea. Specific to System Design, we need to know how to make use of several tools and techniques used for designing. These tools and techniques are Flowchart, Data flow diagram, Data dictionary, Structured English, Decision table, Decision tree.

**Comprehension** is the second level cognition, which is to understand the meaning, translation, interpretation of problems and instructions. The keywords associated with this level are comprehends, explains, infers, interprets and translates. In this level of cognition the programmer having fully comprehending the system, should be able to translate and convert into a design using the tools available for designing. In order to design the system, we should understand the system thoroughly well.

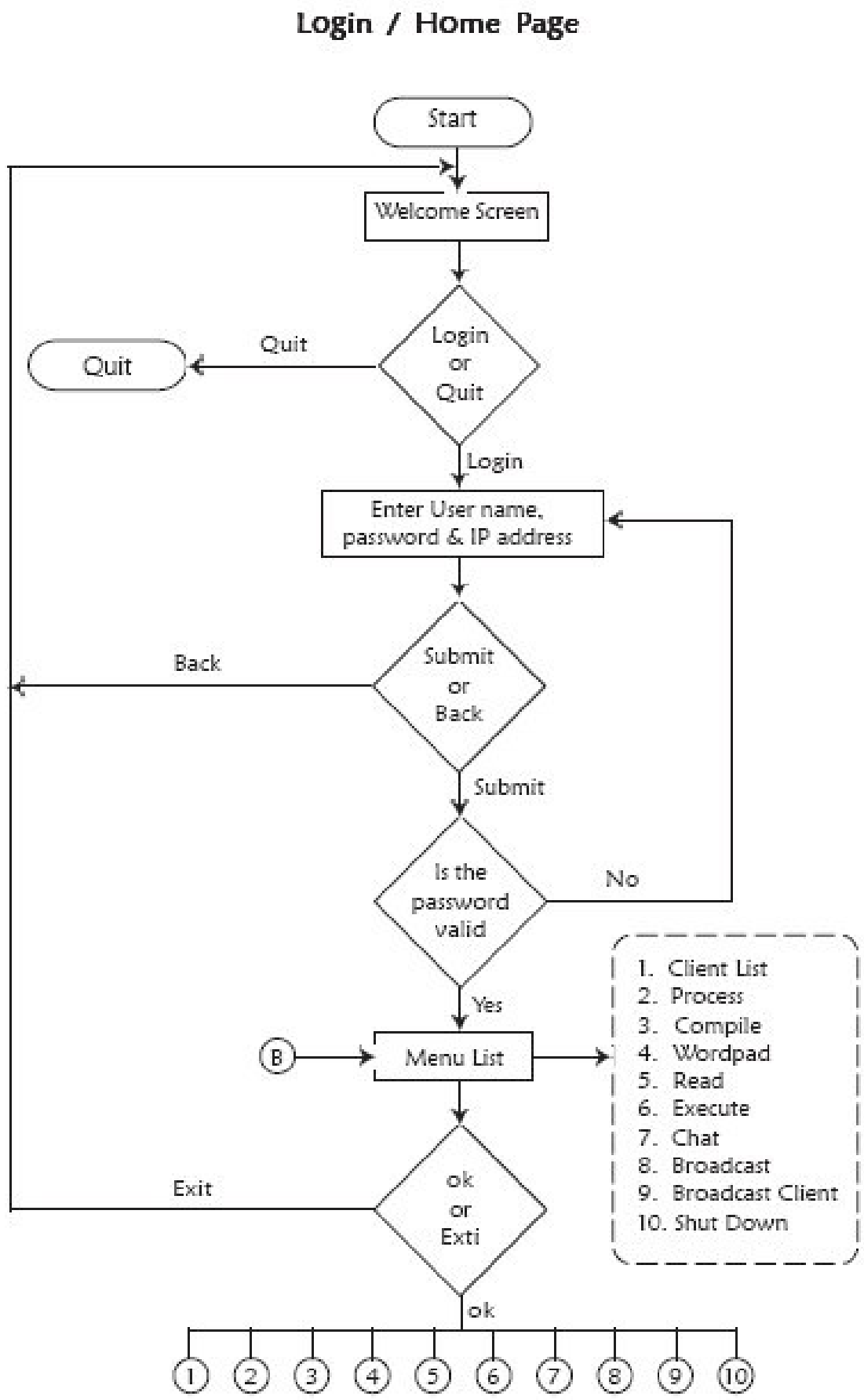

Figure 4 Flowchart-Main menu &Options

**Application** is the third cognition level, which applies the knowledge learned into a novel work. It has the keywords such as relate, transfer, associate, apply, computes, changes, construct, demonstrate, discover, manipulate, prepare, produce and solves. Construction, association and applying flowcharts, pseudo codes, data flow diagrams, block diagrams into the design of software or a system takes place in this cognitive level.

**Analysis** is the fourth in the Bloom's taxonomy which separates concepts into component parts so that its organizational structure may be understood. The keywords in this category are analyzes, compares, breaking down, differentiates, discriminates, distinguishes, identifies, illustrates, selects and separates. In this level we have analyzed and able to distinguishes between the components in our design. We have established the relationship between different components. As we already explained earlier, there are two components in the system, they are one is Mobile component and the other is intranet component. The Intranet component is subdivided into Server part as well as Clients part. The breaking down of tasks took place here. We have separated the concept into different parts so that organizational structure is well distinguished. The different components and the technologies thus separated and distinguished is as shown in the figure 5.

Next cognitive level according to Bloom's Taxonomy is the **Synthesis level**. It builds a structure or pattern from diverse elements together to form a whole with emphasis on creating a new complete structure. The key words used here are combines, compiles, composes, creates, devices, designs, generates, organizes, rearranges, revises and summarizes. There are several instances of synthesis we found in the system design stage.

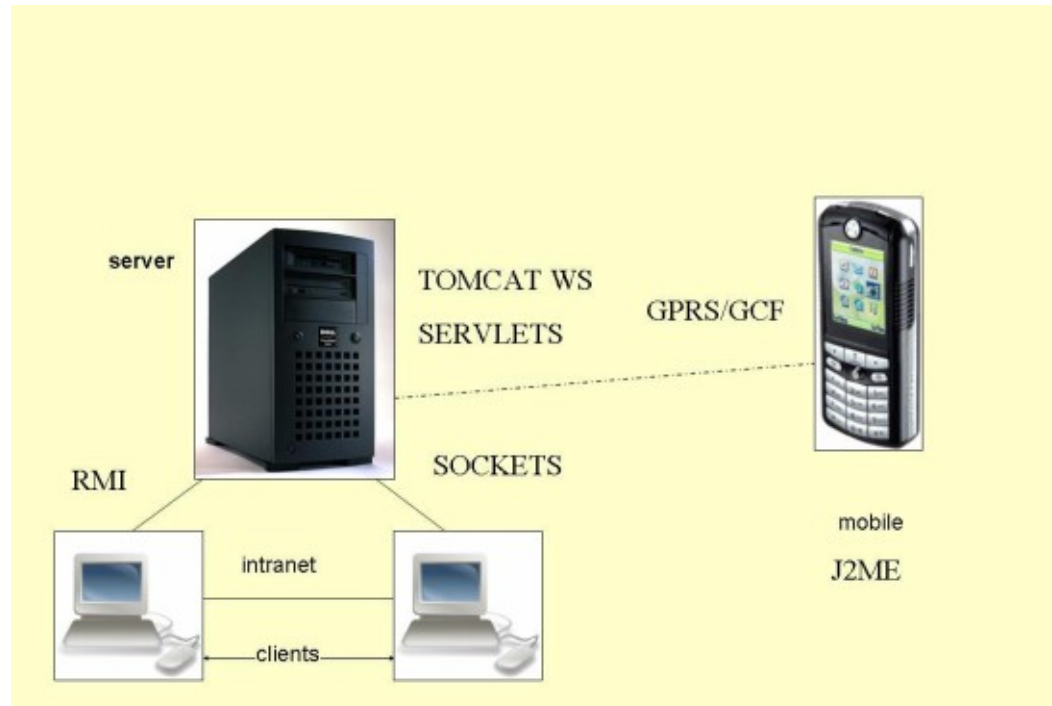

Figure. 5 Components and Technology in each Component

After breaking the tasks into three groups such as Mobile component, server component and clients component. Now we need them to knit together by making of several technologies available. In the design stage itself we needed to identify them. Server part and Clients part anyway they are connected by LAN. The server will be running Tomcat webserver to accept the request from mobile. Mobile and Webserver are connected using GPRS and Generic Connection Framework (GCF) provided in J2ME. When Mobile user trying to establish connection with a client depending upon the request that has to be carried out either Remote Method Invocation (RMI) or socket programming will be activated. Here GPRS combines a mobile user and a server. RMI or Socket will connect a mobile/server and a client.

The last level which is more complex according to Bloom is **Evaluation.** It is about making judgments about the value of ideas or concepts involved in the system. It selects the most effective solution. The keywords or verbs used here are appraises, concludes, contrasts, defends, evaluates, explains, justifies and supports. In this stage, after correlating the technologies and different components in our design, we have evaluated the suitability of them in our scenario. This evaluated hypothesis is validated to justify the decisions already made.

According to the system design study we made it is evident that all our actions carried out in design reflect all the levels of Bloom's taxonomy of hierarchy staring from simple knowledge and comprehension, through application, analysis and finally synthesis and evaluation.

We tried to correlate frequency of activities and each of the Bloom's cognitive level and the result is as tabulated in table 1 and table 2.

The graph shown in the figure 6 summarizes that the system design activity in software engineering involves all the six cognitive levels of Bloom's Taxonomy. It is also evident from the graph that the software design is a critical and complex activity as it highly involves more number of activities in level 5 and level 6 i.e., synthesis and evaluation level respectively.

| sl no | Bloom's Level | Frequency |
|---|---|---|
| | | |
| 1 | Knowledge | 12 |
| | | |
| 2 | Comprehension | 3 |
| | | |
| 3 | Application | 3 |
| | | |
| 4 | Analysis | 3 |
| 5 | Synthesis | 7 |
| | | |
| 6 | Evaluation | 6 |

Table 2: Table of Bloom's Cognitive Levels and Activities in the design

| SL NO | BLOOM'S LEVEL | ACTIVITY |
|---|---|---|
| 1 | Knowledge | Recognize multitier architecture |
| | | Recalling RMI concepts |
| | | Socket programming |
| | | Mobile Programming |
| | | HTTP |
| | | Java Programming |
| | | Identifying Technologies |
| | | Selection of Technologies |
| | | Recalling system design concepts like DFD's, |
| | | Flowcharts, Decision Tables |
| | | Block diagrams |
| | | Pseudo codes |
| 2 | Comprehension | Interpretation of problem i.e., establishing connection between a mobile and a server |
| | | To fully comprehend the system in terms of Remote Monitoring of Intranet from mobile |
| | | Translation of these requirements into a system design |
| 3 | Application | Constructing Flowcharts |
| | | Preparing DFD's |
| | | Applying Pseudo codes |
| 4 | Analysis | Analyzing the design |
| | | Identification the components |
| | | Breaking down into separate components such as Mobile component, Server component and Client component |
| 5 | Synthesis | Building a structure from different components |
| | | Combining different technologies like J2ME |
| | | GPRS |
| | | HTTP |
| | | Tomcat Web server |
| | | Servlets,RMI |
| 6 | Evaluation | Making judgments about suitability and feasibility of each of the following technologies such as J2ME, |
| | | GPRS |
| | | RMI |
| | | Socket Programming |
| | | Servlets |
| | | Selecting most effective solution |
| | | |

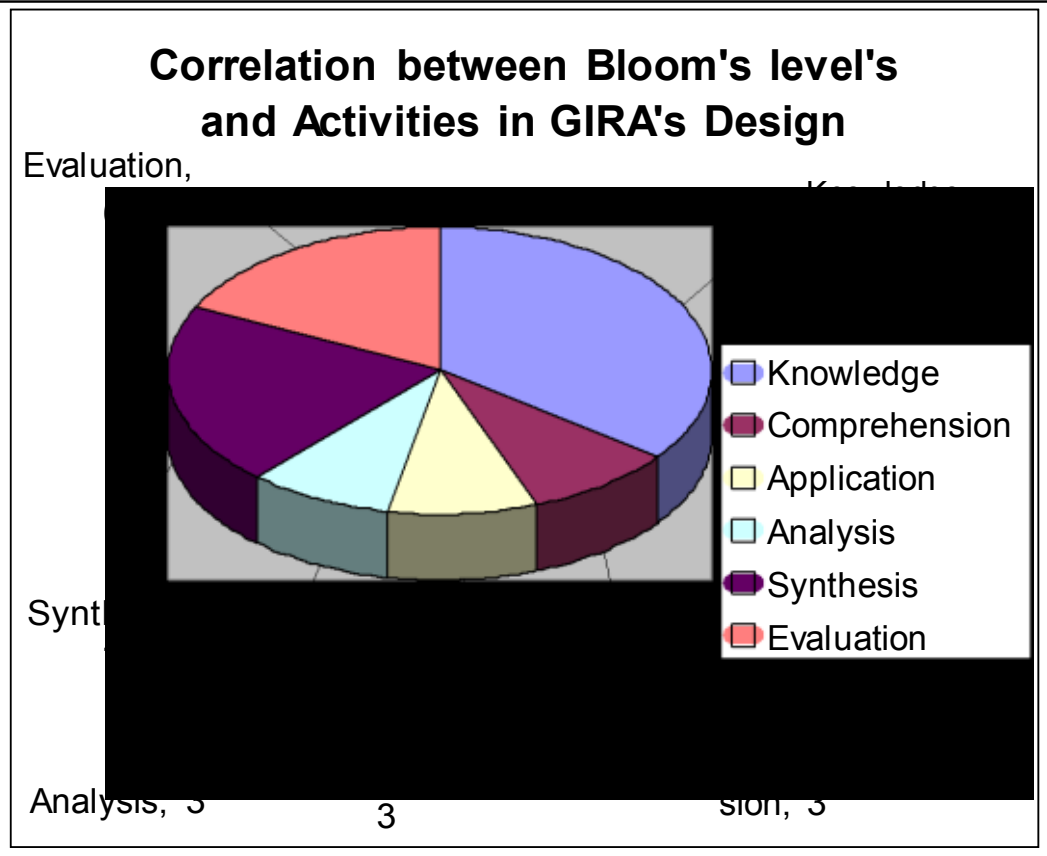

Figure. 6 : Correlation between activities and Bloom's Cognitive Level

## Conclusion and Future Work

We have studied the application of Bloom's taxonomy into the complex program designing aspect. It is evident from the study that all the six levels of Bloom's Taxonomy are used in the complex software design. It is also evident from the observations and statistics made that the software design is a complex activity in software engineering as it uses the complex levels of Bloom's taxonomy such as Synthesis and Evaluation levels.

The case studies for different software design of different software projects may yield different results. Therefore future study may include such case studies. It will help in studying cognitive activities involved in different designs of different software projects. The study may also yield different results depending on the expertise level of the individual programmer involved in the case study. Hence for different set of programmers in varying degree of expertise would also be involved in the future study.